\documentclass[aps,prl,reprint,preprintnumbers,superscriptaddress,amsmath,amssymb,bibnotes,longbibliography]{revtex4-2}

\usepackage{graphicx}
\usepackage{bm}
\usepackage[colorlinks,linkcolor=blue,anchorcolor=blue,citecolor=blue,urlcolor=blue,filecolor=blue,menucolor=blue,runcolor=blue]{hyperref}

\graphicspath{{figures/}}

\begin{document}
	
	
	\title{Pressure-induced strange metal phase in a metallic kagome ferromagnet}
	
	\author{Bin Shen}
	\email{bin.shen@physik.uni-augsburg.de}
	\affiliation  {Experimental Physics VI, Center for Electronic Correlations and Magnetism, University of Augsburg, 86159 Augsburg, Germany}

	\author{Feng Du}
	\affiliation  {Max Planck Institute for Chemistry, Hahn Meitner Weg 1, Mainz 55128, Germany}
	
	\author{Franziska Breitner}
	\affiliation  {Experimental Physics VI, Center for Electronic Correlations and Magnetism, University of Augsburg, 86159 Augsburg, Germany}
	
	\author{Victoria A. Ginga}
	\affiliation{Felix Bloch Institute for Solid-State Physics, University of Leipzig, 04103 Leipzig, Germany}
	
	\author{Ece Uykur}
	\affiliation{Helmholtz-Zentrum Dresden-Rossendorf, Inst Ion Beam Phys \& Mat Res, 01328 Dresden, Germany}
	
	\author{Alexander A. Tsirlin}
	\affiliation{Felix Bloch Institute for Solid-State Physics, University of Leipzig, 04103 Leipzig, Germany}
	
	\author{Philipp Gegenwart}
	\affiliation{Experimental Physics VI, Center for Electronic Correlations and Magnetism, University of Augsburg, 86159 Augsburg, Germany}
	
	\date{\today}

	\begin{abstract}
		Strange metallicity with $T$-linear electrical resistance preceding high-$T_c$ superconductivity remains an enigmatic, yet crucial, signature of correlation physics. Using electrical transport and magnetization measurements up to 50~GPa, we show that such a strange-metal phase is formed in pressurized kagome ferromagnet CrNiAs. In contrast to other kagome materials, a linear suppression of the Curie temperature is found, with the ferromagnetic quantum critical point at $p_{\rm{c}} \approx 12.5$~GPa. Remarkably, from $p_{\rm{c}}$ up to the highest measured pressure, characteristic strange-metal behavior is observed, whereas magnetic field reinstates the Fermi liquid.
		Electronic structure calculations reveal robust weakly dispersive bands persisting unchanged beyond $p_{\rm{c}}$, possibly at the origin of the $T$-linear electrical resistance. This establishes pressurized kagome ferromagnets as an intriguing platform for strange-metal behavior.
		
	\end{abstract}
	
	\maketitle
	
	While electronic properties of ordinary metals are well described in the framework of Landau's Fermi-liquid (FL) theory, ``strange-metal'' behavior with the non-Fermi liquid (NFL) $T$-linear electrical resistivity 
	from high to very low temperatures has been associated with a breakdown of the quasiparticle description \cite{08GenNP, 19LegNP, 22PhiSci}. The theoretical understanding of this behavior remains one of the central problems in the field of correlated electrons. The underlying quantum critical points (QCPs) were explored in strongly correlated materials, such as heavy-fermion metals and cuprate or pnictide superconductors. More recently, a related behavior is discussed in connection to flat bands in moiré structures and frustrated lattice compounds~\cite{22JaoNP,24CheNRM}.
	
	
	While antiferromagnetic quantum criticality is broadly observed and theoretically well established, a distinct behavior is found for the suppression of ferromagnetic (FM) order. Theoretically, a FM QCP appears to be naturally avoided in itinerant ferromagnets \cite{99BelPRL, 04ChuPRL, 16BraRMP} by either an abrupt suppression of ferromagnetism, a change from ferromagnetic to antiferromagnetic order, or, in the case of significant disorder, by a smeared quantum phase transition~\cite{16BraRMP}. In fact, only a few FM Kondo lattices, such as YbNi$_4$P$_2$~\cite{13AleSci} and CeRh$_6$Ge$_4$~\cite{20SheNature}, display a FM QCP. In the vicinity of the critical substitution or critical pressure, these systems reveal strange-metal behavior \cite{13AleSci, 20SheNature}. Various theories put forward the essential ingredients of FM quantum criticality, such as persistent local moments \cite{20SheNature}, a non-centrosymmetric crystal structure \cite{20KirPRL}, a certain degree of magnetic anisotropy \cite{22WanSCMA}, or nonsymmorphicity \cite{24ShiNC}. However, these theories have yet to be subjected to the rigors of experimental testing.
	
	\begin{figure*}
		\includegraphics[angle=0,width=0.98\textwidth]{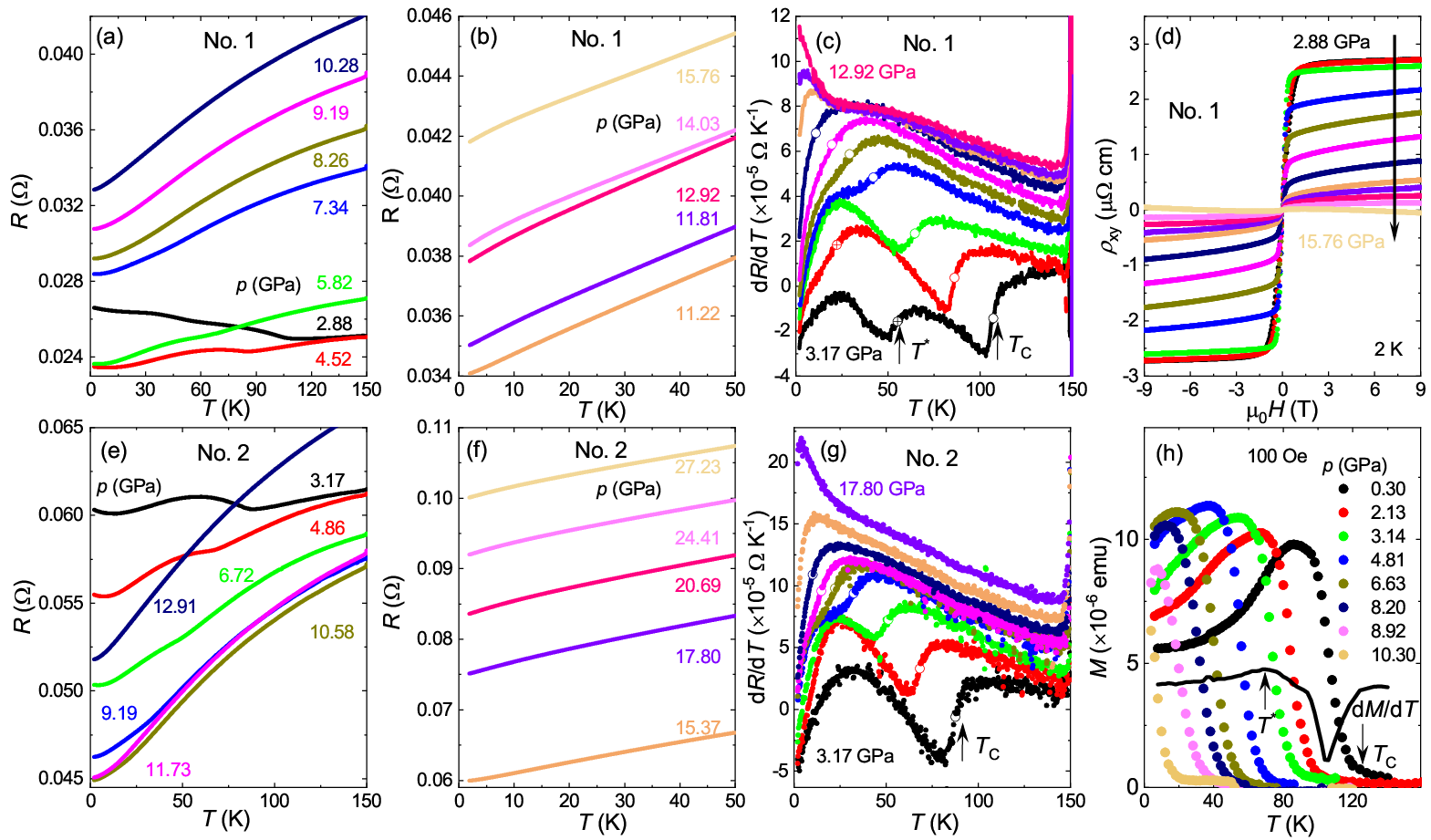}
		\vspace{-12pt} \caption{\label{Fig1} (a-b) Temperature-dependent resistance of CrNiAs and (c) its derivative for run No. 1. (d) Hall resistivity measured at 2~K for run No. 1. (e-f) Resistance of CrNiAs and (g) its derivative for run No. 2. The arrows together with the open symbols in (c) mark the transition temperatures $T_{\rm{C}}$ and $T^{\star}$, respectively, while in (g) for run No. 2, arrows and open symbols indicate the transition temperature $T_{\rm{C}}$. (h) Temperature-dependent dc magnetization of CrNiAs in a magnetic field of 100~Oe. The solid line shows the d$M$/d$T$ at 0.3~GPa. The arrows label the transition temperatures $T_{\rm{C}}$ and $T^{\star}$, respectively.}
		\vspace{-12pt}
	\end{figure*}
	
	Kagome materials offer a new playground for the FM QCP and strange metallicity \cite{24YeNP, 24LiuN}. Generically, the geometrically frustrated two-dimensional kagome lattice can host non-trivial topological electronic states associated with the Dirac points, van Hove singularities, and flat bands \cite{13WanPRB, 13KiePRL}. These features result in a variety of quantum phases, including magnetism, charge ordering, and superconductivity \cite{22YinN, 23WanN, 24SteNature}. While pressure tuning has successfully been used to study NFL behavior at the suppression of the density wave in CsCr$_3$Sb$_5$~\cite{24LiuN}, tuning FM kagome metals across their quantum phase transition has not yet been demonstrated.
	
	We focus on CrNiAs, crystallizing in a noncentrosymmetric hexagonal structure (space group, P$\overline{6}$2m), with a distorted kagome lattice of magnetic Cr atoms. From polycrystals it is known that CrNiAs undergoes a FM transition with $T_{\rm C}$ ranging from 170 to 190~K, followed by another transition $T^\star$ at 110~K \cite{95SatJMMM, 08StaJPCM, bacmann2004}. It has also been shown that both $T_{\rm C}$ and $T^{\star}$ are susceptible to alterations in the Cr:Ni stoichiometry \cite{80IwaJPSJ}.  Neutron scattering and $^{\rm 61}$Ni M$\ddot{\rm o}$ssbauer measurements indicate that Cr atoms are responsible for the ferromagnetism in this material \cite{bacmann2004, 08StaJPCM}, which is also consistent with electronic structure calculations \cite{96IshPB, 97TobJAC}. A decrease of $T_{\rm{C}}$ and $T^{\star}$ has been observed in polycrystalline samples subjected to hydrostatic pressure up to 1.3~GPa \cite{95SatJMMM}. 
	
	Below, we present the results of electrical transport and magnetization measurements conducted under hydrostatic pressure on single crystalline CrNiAs. The complete pressure-temperature phase diagram is determined. We uncover the linear decrease in $T_{\rm{C}}$ with the eventual suppression of the FM phase at a critical pressure $p_{\rm{c}}$ $\approx$ 12.5~GPa. Most remarkably, above $p_{\rm{c}}$ and up to almost 50~GPa, within the nonmagnetic phase of CrNiAs, a robust strange-metal phase with the $T$-linear resistivity is observed in zero magnetic field, yet it turns into the conventional $T^2$ FL behavior at low temperatures when magnetic field is applied. 
	
	Single crystals were grown by the Bi-flux method. The samples used in this study were characterized by measurements of magnetization, electrical transport, and heat capacity (see Fig. S1). Our samples show $T_{\rm C}$ = 135~K and $T^\star$ = 90~K. The resistance and Hall resistivity were measured in a diamond anvil cell (DAC) utilizing the PPMS (Quantum Design). The magnetization under pressure was measured in the ceramic anvil cell with the MPMS3 SQUID magnetometer (Quantum Design). The electronic structure was calculated by the density functional theory (DFT). Further details on the sample synthesis, measurements, and calculations are described in the Supplementary Materials \cite{CrNiAsSI}. Importantly, the results were reproduced in three runs of electrical transport measurements using different pressure-transmitting media and diamond anvils with different culet sizes. 
	
	At ambient pressure, the electrical resistance $R(T)$ of CrNiAs monotonically increases upon cooling from 300~K [see Fig.~S1(c)], while already at the lowest measured pressure of 2.88~GPa a metallic temperature dependence is found in the paramagnetic state, as shown in Figure~\ref{Fig1} (a). At low pressures, $R(T)$ starts increasing on cooling the system through $T_{\rm{C}}$ and $T^{\star}$, indicating that additional scattering mechanisms appear below these transitions. The transition temperatures were determined by analyzing the temperature derivative of the resistance [Fig.~\ref{Fig1} (c)]. At higher pressure, both transitions are gradually suppressed [Fig.~\ref{Fig1} (c)], and metallic behavior of the resistance across the whole temperature range is observed already around 6~GPa where $T^{\star}$ disappears. The critical pressure for $T_{\rm{C}}$ is found to be approximately $p_{\rm{c}}=12$~GPa. At the pressure of 11.22~GPa [Fig.~\ref{Fig1} (b)], the resistance shows a quasi-linear temperature dependence from 2~K up to at least 50~K (note the weak variation of $dR/dT$ with temperature at this pressure shown in panels (c)). Upon further increasing pressure, this quasi-linear temperature dependence is maintained. The above behavior has been consistently reproduced in two subsequent runs [Fig.~\ref{Fig1} (e-g) for No. 2, Fig. S2, and S3 in supplement for No. 3 up to the highest pressure of almost 50~GPa]. Note that such a wide pressure regime of the low-temperature NFL behavior is incompatible with ordinary quantum criticality and suggests the formation of an extended strange-metal phase.
	
\begin{figure}
		\includegraphics[angle=0,width=0.42\textwidth]{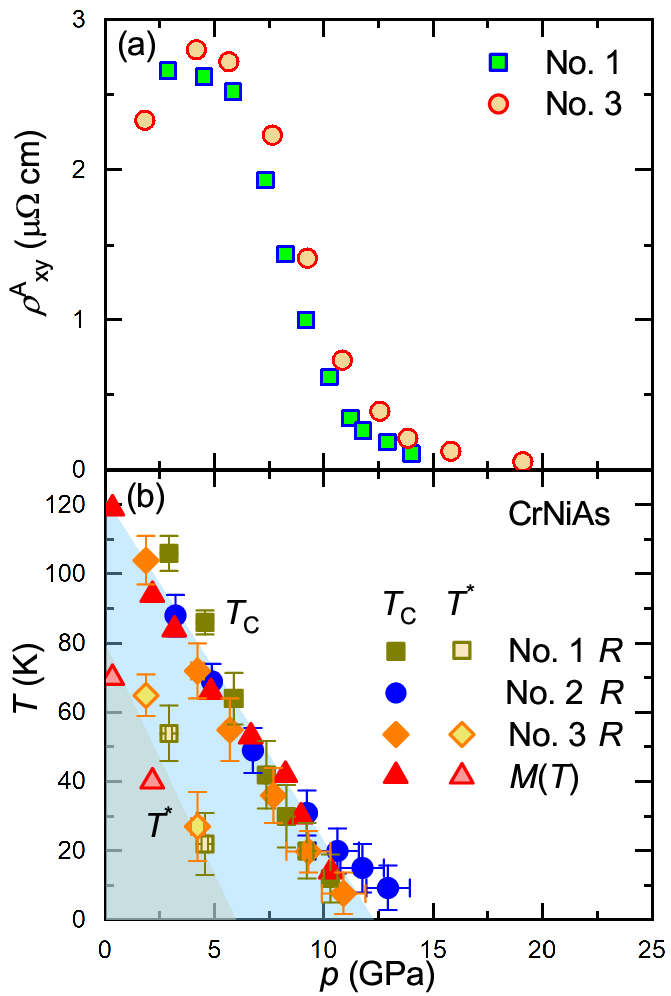}
		\vspace{-12pt} \caption{\label{Fig2} (a) Evolution of the anomalous Hall resistivity $\rho_{xy}^A$ as a function of pressure, extracted from run No. 1 and 3. (b) Pressure-temperature phase diagram of CrNiAs inferred from resistance of three runs and magnetization measurement. $T_{\rm{C}}$ is roughly linearly suppressed by hydrostatic pressure with the critical pressure $p_{\rm{c}}$ $\approx$ 12.5~GPa.}
		\vspace{-12pt}
	\end{figure}

	To follow the FM hysteresis with pressure, we also measured for run No. 1 the Hall resistivity at 2~K, as shown in  Fig.~\ref{Fig1} (d). Within the FM state the anomalous Hall effect can be clearly detected. Importantly, the anomalous Hall signal decreases {\it gradually} with increasing pressure.  Above 14~GPa, no anomalous Hall effect is apparent. We describe the Hall resistivity in the FM state by the conventional formula $\rho_{xy}$ = $\rho_{xy}^0$ + $\rho_{xy}^A$ = $R_0 B$ + $R_s \mu_{0} M$, where $\rho_{xy}^0$ is the normal part, and $\rho_{xy}^A$ denotes the anomalous contribution. $R_0$ and $R_s$ are the corresponding normal and anomalous Hall resistivity coefficients, respectively. The fitting is performed from 5~T to 9~T, as the Hall resistivity shows linear field dependence in this window. Similar measurements and analyses were conducted on run No.~3 (Fig. S4). The extracted pressure-dependent $\rho_{xy}^A$ is displayed in Fig.~\ref{Fig2} (a) for both runs. Below 6~GPa, $\rho_{xy}^A$ shows a weak pressure dependence. Upon further compression, $\rho_{xy}^A$ is gradually reduced. Above 12.5~GPa, the $\rho_{xy}^A$ value becomes significantly reduced, implying the loss of ferromagnetism.
	
	Figure~\ref{Fig1} (h) shows the magnetization $M(T)$ under pressure up to 10.30~GPa, which represents the maximum pressure that can be applied in such magnetometer cells.  $T_{\rm{C}}$ was determined as the onset temperature of a sudden increase in the magnetization. At low pressures (0.30 and 2.13~GPa), $T^{\star}$ can also be observed as a kink in $M(T)$, together with a peak in d$M$/d$T$. With increasing pressure, $T_{\rm{C}}$ shifts towards lower temperatures. Even at the highest applied pressure of 10.30~GPa, the magnetic behavior remains FM, with $T_{\rm{C}}$ $\approx$ 18~K.        
	
	The pressure-temperature phase diagram of CrNiAs from the electrical transport and magnetization measurements is shown in Fig.~\ref{Fig2} (b). The data indicate a gradual suppression of $T^{\star}$ under pressure, with a critical pressure of 6~GPa. Furthermore, the Curie temperature is also found to be continuously suppressed. This indicates a QCP around a critical pressure of $p_{\rm{c}}$ $\approx$ 12.5~GPa at which the ferromagnetism is completely destroyed. Above $p_{\rm{c}}$, strange-metal behavior with a quasi-linear $T$-dependent resistance is observed at low temperatures, extending up to $\sim 50$~GPa [see Fig.~\ref{Fig1} (b) and (f) and Fig. S3 (b)].
	
	\begin{figure*}
		\includegraphics[angle=0,width=0.99\textwidth]{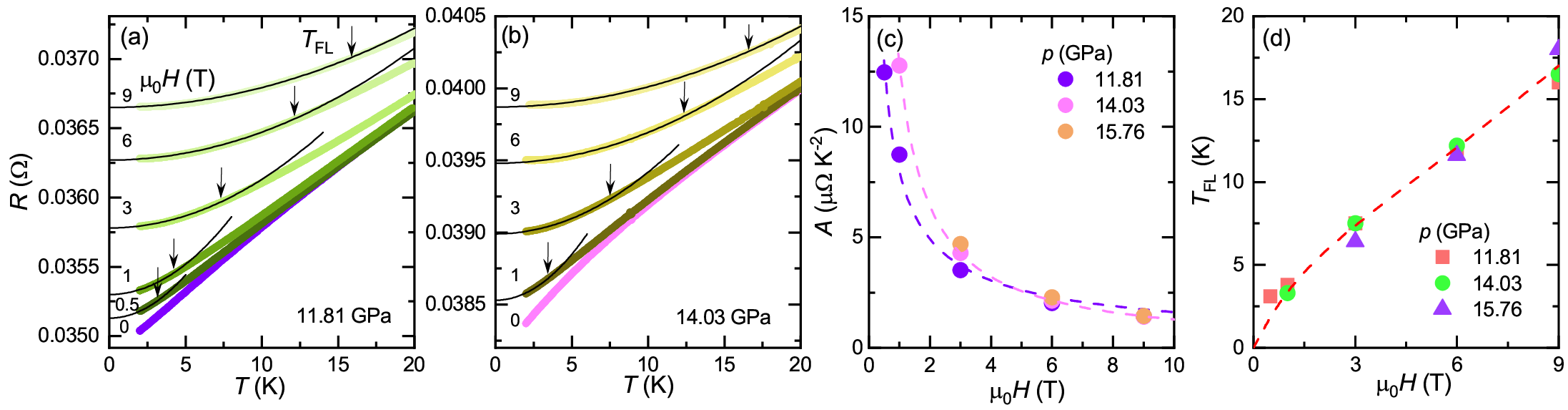}
		\vspace{-12pt} \caption{\label{Fig3} Temperature-dependent resistance in various magnetic fields at (a) 11.82 and (b) 14.03~GPa. Black solid lines are the Fermi liquid $AT^2$ fits and arrows mark the Fermi liquid temperature $T_{\rm{FL}}$. (c) Fermi-liquid $A$-coefficient as a function of the magnetic field at 11.82, 14.03 and 15.76~GPa. The error bars for the $A$-coefficient are smaller than the symbols. The dashed lines show a $H^{\rm{m}}$ fit to the data at 11.82 ($m = -0.7$) and 15.76~GPa ($m = -1.0$), respectively. (d) $T_{\rm{FL}}$ as a function of magnetic field at different pressures.}
		\vspace{-12pt}
	\end{figure*}
	
	We now turn to the influence of magnetic fields on the strange-metal behavior. Figure~\ref{Fig3} shows the resistance $R(T)$ in various magnetic fields below 20 K. At 11.82~GPa [Fig.~\ref{Fig3} (a)], the low-temperature linear temperature dependence is gradually replaced by a Fermi-liquid-like quadratic temperature dependence as magnetic field is increased. This suggests that the strange-metal behavior arises from spin fluctuations that are suppressed by field. Furthermore, resistance increases with field, indicating an overall positive magnetoresistance. A similar behavior is also observed at pressures of 14.03~GPa [Fig.~\ref{Fig3} (b)] and 15.76~GPa (Fig. S5). By fitting the low-temperature resistance with the formula $R(T) = R_0 + A T^2$, where $R_0$ is the residual resistance and $A$ stands for the Fermi-liquid coefficient, we obtain the magnetic field dependence, $A(H)$, plotted in Fig.~\ref{Fig3} (c). It shows a power-law divergence $H^{\rm{m}}$, with $m$ being -0.7 (-1.0) at 11.82 (14.03)~GPa towards zero field, indicative of zero-field quantum criticality. Fig.~\ref{Fig3} (d) displays the temperatures $T_{\rm{FL}}$ below which the FL behavior is realized. This behavior indicates a suppression of the field-induced FL in the zero-field limit, compatible with a quantum critical state.
	
	\begin{figure}
		\includegraphics[angle=0,width=0.48\textwidth]{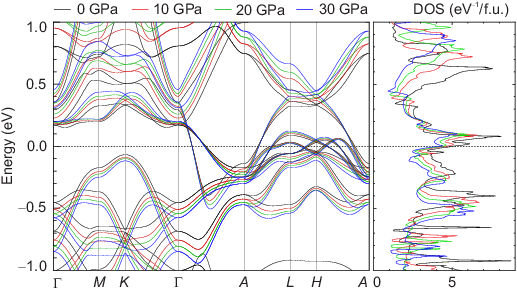}
		\vspace{-12pt} \caption{\label{DFT} Pressure dependence of the band structure and density of states for nonmagnetic CrNiAs. The Fermi level is at zero energy.}
		\vspace{-12pt}
	\end{figure}   
	
	\textit{Discussion:} Prior to examining the properties of CrNiAs under pressure, we comment on the lower $T_{\rm{C}}$ and $T^\star$ of our single crystals compared to the previous literature~\cite{95SatJMMM, 08StaJPCM, bacmann2004,80IwaJPSJ}. Iwata \textit{et al.}~\cite{80IwaJPSJ} reported single crystal growth by the thermal annealing method and showed the Curie temperature of $T_{\rm{C}}$ = 190~K in stoichiometric CrNiAs. Both $T_{\rm{C}}$ and the saturation moment are reduced if $x$ deviates from 1 in Cr$_{\rm{1-x}}$Ni$_{\rm{1+x}}$As. Using the results of Ref.~\cite{80IwaJPSJ}, we conclude that our crystals may exhibit a slight Ni excess with $x \approx$ 0.08. This deviation from the ideal stoichiometry could potentially induce disorder and enhance the resistivity. However, the thermal annealing method employed in Ref.~\cite{80IwaJPSJ} typically results in an As deficiency due to the low evaporation temperature of As, whereas the flux method adopted here can greatly mitigate such problems. The resistivity of our single crystals is one order of magnitude lower than that of polycrystalline samples \cite{95SatJMMM}, indicating the high quality of our samples. Normally, disorder-induced NFL behavior is manifested by the robustness of the electrical transport in magnetic fields. However, this is not the case here, as illustrated in Fig.~\ref{Fig3}. This finding suggests that the observed strange-metal behavior in CrNiAs is intrinsic, rather than being caused by residual structural disorder.                
	
	Numerous FM metals have been observed to exhibit a discontinuous suppression of magnetic order under pressure, such as CoS$_2$ \cite{97GotPRB}, ZrZn$_2$ \cite{04UhlPRL}, and UGe$_2$ \cite{10TaoPRL}. In this context, it is remarkable that we can follow the clear signatures of $T_{\rm{C}}$ in CrNiAs from 135~K at ambient pressure down to 7~K (which corresponds to only 5~\% of the original value) in the vicinity of the critical pressure [Fig.~\ref{Fig2} (b)]. Our data do not show the abrupt disappearance of $T_{\rm{C}}$ and indicate a continuous suppression of ferromagnetism. This conclusion is further supported by the smooth decrease of the anomalous Hall signal when approaching $p_{\rm{c}}$, as $\rho_{xy}^A$ $\propto$ $M$ [Fig.~\ref{Fig1} (d) and Fig.~\ref{Fig2} (a)].  
	
	Near the critical pressure, the residual resistivity is estimated to be approximately 350~$\mu\Omega$ cm, corresponding to the residual resistivity ratio $RRR$ (here defined as $R_{\rm{300~K}}/R_{\rm{2~K}}$) of 1.4. However, this does not place CrNiAs in the category of disordered systems. It is important to note that other compounds with the same crystal structure, such as CrRhAs \cite{95SatJMMM} and FeCrAs \cite{09WuEPL, 14AkrPRB}, also exhibit large residual resistivities and even a nonmetallic temperature dependence. In these compounds, Cr atoms are magnetic, while the other transition-metal atoms form a nonmagnetic sublattice. The anomalous electrical transport behavior is also robust under hydrostatic pressure \cite{13TafJPCM}, and the high resistivity appears to be intrinsic to these compounds with a similar structural motif. While magnetic Cr atoms form a distorted kagome structure, every three Ni (or Fe) atoms form a trimer, which are then connected in the form of triangles \cite{15FloJPCM}. It is possible that this particular structure may induce frustration, which could subsequently enhance scattering of the carriers. The nonmagnetic nature of the Ni (or Fe) sublattice provides evidence for the presence of frustration. An intriguing theory has been put forth that in FeCrAs, the nonmagnetic Fe sublattice serves as a host for critical fluctuations due to its vicinity to a metal-insulator transition \cite{11RauPRB}. Whether the Ni sublattice in CrNiAs exhibits comparable physics still remains to be investigated. An additional potential cause of the elevated resistivity observed in these compounds is the high-energy spin fluctuations \cite{18PluPRB}. On the other hand, disorder generally enhances the resistivity power $n$ in the vicinity of a quantum phase transition \cite{00GroLPCM}. In particular, in ferromagnets, the disorder effect gives rise to $n = 3/2$ \cite{04KhaPRB, 18KirPRB}. Thus, the observed quasi-linear temperature-dependent resistance in CrNiAs is unlikely to be attributed to the disorder effect.    
	
	It is noteworthy that analogous extended NFL behavior has been observed in several FM metals upon suppressing their magnetism by pressure, as in ZrZn$_2$ \cite{07TakJPSJ}, Ni$_3$Al \cite{05NikPRB}, and the helimagnet MnSi \cite{01PflN, 03DoiN}. However, the power $n$ in these cases ranges from 3/2 to 5/3, which is very different from 1 in our case. A marginal Fermi liquid model of enhanced spin fluctuations has been proposed to explain the NFL behavior in ZrZn$_2$ \cite{08SmiNaure, 12SutPRB} and Ni$_3$Al \cite{05NikPRB}. An alternative scenario involves antiferromagnetic short-range fluctuations due to the Fermi surface nesting \cite{12KabJPSJ}. In the case of MnSi, the NFL behavior is attributed to a phase inhomogeneity or partial long-range order \cite{04YuPRL, 04PflNaure}, rather than quantum criticality \cite{07PflSci, 24DalPRL}. Nevertheless none of the aforementioned interpretations can be applied to the quasi-linear in $T$ resistance observed over a wide pressure range in CrNiAs.  
	
	The persistence of the $T$-linear resistivity at pressures that are 3-4 times higher than $p_{\rm c}$ suggests that this strange-metal behavior is not caused by a mere proximity to the ferromagnetic QCP. Indeed, our DFT calculations reveal weakly dispersive bands near $E_{\rm F}$ along the $L-H$ reciprocal direction (Fig.~\ref{DFT}). Whereas all other bands show a visible broadening upon compression, these weakly dispersive bands remain almost unchanged and pinned to the Fermi level up to at least 30~GPa, indicating its importance for the strange-metal phase. They are reminiscent of flat bands observed in other kagome metals, such as Ni$_3$In \cite{24YeNP} and CoSn \cite{20KangNC}. 
	
	It is interesting to compare CrNiAs to the quasi-one-dimensional heavy-fermion compound $\beta$-YbAlB$_4$ where an antiferromagnetic order is induced beyond the critical pressure of 2.5~GPa. From ambient pressure up to $p_{\rm{c}}$
	the electrical resistance exhibits a linear temperature dependence, $\Delta \rho \propto T$ \cite{15TakSci, 16TomPRB}. Moreover, magnetic field destroys the NFL behavior in $\beta$-YbAlB$_4$ and induces FL behavior in resistivity with a divergent coefficient $A$ when approaching zero field \cite{08NakNP}. This behavior bears a striking resemblance to our observations on CrNiAs, with the exception of the different nature of magnetism. In the case of $\beta$-YbAlB$_4$, slow critical charge fluctuations possibly associated with the NFL behavior have been detected \cite{23KobSci}. Similarly, charge fluctuations have been proposed to occur on the Fe sublattice in FeCrAs \cite{11RauPRB}, which may be of relevance to the Ni counterpart in CrNiAs.               
	
	In summary, our study reports two important aspects of quantum phase transitions in ferromagnetic metals. First, we have identified CrNiAs as a kagome metal where FM order can be continuously suppressed by hydrostatic pressure. Second, we have uncovered a uniquely broad pressure range of the strange-metal behavior above the critical pressure, in stark contrast to canonical quantum criticality. There are several aspects of our study that merit further investigation, such as the role of the nonmagnetic Ni sublattice and the influence of $T^{\star}$ on ferromagnetism. In order to understand these exotic behaviors, chemical substitution may be a viable approach considering the experimental limitations imposed by the high critical pressure (around 12.5~GPa) for the pristine sample.
	
	The experimental data associated with this manuscript are available from Ref.~\cite{CNADATA}.
	
	\section{Acknowledgments}
	This work was funded by the Deutsche Forschungsgemeinschaft (DFG, German Research Foundation) – TRR 360 – 492547816. B.S. was supported by the Alexander von Humboldt Foundation. We thank Jihaan Ebad-Allah, Anton Jesche, Christine Kuntscher and Mikhail Ivanovich Eremets for valuable discussions.

\end{document}